\documentstyle[11pt]{article} 
\normalbaselines
\begin{document} 
\bibliographystyle{unsrt}

\vbox{\vspace{38mm}} 
\begin{center} 
{\LARGE \bf A new scheme for the Klein-Gordon and Dirac \\[2mm] 
fields on the lattice with axial anomaly}\\[5mm]

Miguel Lorente\\{\it Departamento de F\'{\i}sica, Universidad de Oviedo, 
33007 Oviedo, Spain\\Arnold Sommerfeld Institut f\"{u}r Theoretische Physik, 
Technische Universit\"{a}}t, D-3392 Clausthal, Germany
\\[5mm] 
(Submitted 30 October 1992)\\[5mm] \end{center}

\begin{abstract} Using the method of finite differences a scheme is proposed 
to solve exactly the Klein-Gordon and Dirac free field equations, in a 
$(1+1)$-dimensional lattice. The hamiltonian of the Dirac field is
translational invariant, hermitian, avoids fermion doubling, and, for the
massless case, preserves global chiral symmetry. Coupling the fermion field to
the electromagnetic vector potential we construct a gauge invariant vector 
current leading to the correct axial anomaly. PACS number: 02.20.+b, 11.30.-j
\end{abstract}

\section{Introduction} 
The method of finite elements has become very powerful to solve evolution
equations in quantum field theories. In particular Bender and his
collaborators \cite{Be} have applied the method to the Klein-Gordon and Dirac
quantum field equations, consistent with the equal time commutation relations,
which preserves global chiral symmetries in the massless case and avoids
fermion doubling. Matsuyama \cite{Mat} has constructed an abelian gauge-coupled
Dirac equation without species doubling. Milton \cite{Mil} extended this method
to the non-abelian gauge theories. Using the method of finite differences,
V\'{a}zquez \cite{Vaz} also introduced an explicit scheme for the Dirac fields,
such that the associated action is hermitian, the chiral symmetry is preserved
and the fermion doubling is absent, but the equal time commutation relations
are not satisfied.Nevertheless in all these models the solution is given for
only one time step in the transfer matrix. 

Lattice field theories have become very popular to analyze perturbation
theories in the standard model \cite{Smit}. With the lattice regularization
comes the problem of fermion doubling if one imposes to the hamiltonian
translational invariance, hermiticity and locality. Under these assumptions
fermion doubling is unavoidable, as Nielsen and Ninomiya have proved
\cite{Niel}. 

We present a simple model for the Klein-Gordon and Dirac free field using the
method of finite differences. The novelty of our approach lies, first, {\sl on
the exact solutions for any time step} of the difference equations, making a
good contact with the covariant continuous theory (we have already worked out
exact solutions for the Heisenberg difference equations of motions in a time
lattice \cite{Lor}). Secondly, we try to escape the no-go theorem of Nielsen
and Ninomiya because our Hamiltonian is non-local, although avoids fermion
doubling. 

To check our model we apply it to current algebra, namely, we implement the
vector and axial current of fermion massless fields with abelian gauge fields.
The axial current is then gauge invariant, but exhibit non-local point
separation structure, which in the continuous limit leads to the correct axial
anomaly.

\section{A set of orthonormal functions on the lattice} 
Let us define the function
\begin{equation}
f_j={\left({{1+{1 \over 2}ik\varepsilon \over 1-{1 \over 2}ik\varepsilon
}}\right)}^j \quad , \qquad j =0,\pm 1,\pm 2,\ldots 
\label{2.1}\end{equation}
where $k$ is an arbitrary constant and $\varepsilon$ some small quantity.

This function satisfies the difference equations
\begin{eqnarray}
f_{j+1}-f_j &=& ik \varepsilon {1 \over 2}\left({f_{j+1}+f_j}\right) \qquad ,
\quad f_j-f_{j-1} = ik \varepsilon {1 \over 2}\left({f_j+f_{j-1}}\right)
 \label{2.2}\\
f_{j+{1 \over 2}}-f_{j-{1 \over 2}} &= &ik\varepsilon{1 \over 2}\left(
{f_{j+{1 \over 2}}+f_{j-{1 \over 2}}}\right)
\label{2.3}\end{eqnarray}

Using the notation for the difference operators
$$ \Delta {f}_{j} \equiv {f}_{j +1}-{f}_{j}
\quad , \quad \nabla {f}_{j} ={f}_{j} -{f}_{j -1}
\quad , \quad \delta {f}_{j} \equiv {f}_{j +{1 \over
2}}-{f}_{j -{1 \over 2}}$$ 
and the average operators
$$\tilde{\Delta}{f}_{j} \equiv {1 \over 2}\left({{f}_{j
+1}+{f}_{j}}\right) \quad , \quad \tilde{\nabla}{f}_{j} \equiv {1 \over
2}\left({{f}_{j} +{f}_{j -1}}\right) \quad , \quad \mu {f}_{
j} \equiv {1 \over 2}\left({{f}_{j +{1 \over 2}}+{f}_{j -{1
\over 2}}}\right)$$
the above equations become
\begin{eqnarray}
{1 \over \varepsilon}\Delta {f}_{j} &=& i k
\tilde{\Delta}{f}_{j} ,{1 \over \varepsilon}\nabla {f}_{j}
= i\mu \tilde{\nabla}{f}_{j} \label{2.2'}\\
{1 \over \varepsilon} \delta {f}_{j} &=& i 
k \mu {f}_{j}
\label{2.3'}\end{eqnarray}

These difference equations and their solutions are approximation of the
differencial equation
\begin{equation}
\frac{{\rm d}f}{{\rm d}x} = ikf 
\label{2.4}
\end{equation}
with truncation error of second order in $\varepsilon$ (the truncation error is
the difference between the Taylor expansion of the finite difference equation
and the differential equation).

On the other hand the function $f_j$ converges to the solution of the
differential equation (\ref{2.4}):
\begin{equation}
{f}_{j}={{\left({{1+{1 \over 2}ik{x \over j} \over 1-{1 \over 2}ik
{x \over j}}}\right)}^{j}} \quad \rightarrow \quad {e}^{ikx}=f\left({x}\right)
\label{2.5}
\end{equation} 
when $j \rightarrow \infty , \quad j \varepsilon \rightarrow
 x , \quad \varepsilon \rightarrow 0$ with order of
convergence $O\left(\varepsilon^2\right).$

If we impose the boundary conditions
$f_0=f_N=1$ for some fixed number $N$, we find $N$ different values of $k$
\begin{equation}
k_m = \frac{2}{\varepsilon}\tan \frac{\pi m}{N}\quad , \quad m=0,1,\ldots,N-1
\label{2.6}
\end{equation} 
with properties $k_{m+N}=k_m$ and $k_{-m}=-k_m$. Therefore
\begin{equation}
{f}_{j} \left({{k}_{m}}\right) ={\left({{1+{1 \over 2}
i \varepsilon {k}_{m} \over 1-{1 \over 2} i 
\varepsilon {k}_{m}}}\right)}^{j} , \qquad m
=0,1,\ldots , N -1
\label{2.7}
\end{equation}

Notice that if $m=N/2$ for $N$ even, then $f_j\left(k_m\right)
=\left(-1\right)^j$ . Obviously, 
\begin{equation} {f}_{j} \left({{k}_{m}}\right)
=exp\left({i {2 \pi \over N} mj}\right)
\label{2.8}
\end{equation}

These functions satisfy the orthogonality relations
\begin{equation}
{1 \over N} \sum\nolimits\limits_{j =0}^{N -1}
{f}_{j}^{*}\left({{k}_{m}}\right) {f}_{j}
\left({{k}_{m'}}\right)={\delta}_{mm'}
\label{2.9}
\end{equation}
with respect to the scalar product in the Hilbert space
$\not\subset\left(\left[0,N\right]\right)$. With the help of this orthogonality
relation we can express any function $F_j$ of discrete variable in terms of the
$N$ orthogonal functions $f_j\left(k_m\right)$ , namely,
\begin{equation}
{F}_{j} ={1 \over \sqrt {N}} \sum\nolimits\limits_{
m =0}^{N -1} {a}_{m}{f}_{j} \left({{
k}_{m}}\right) 
\label{2.10}
\end{equation}
with
\begin{equation}
{a}_{m} ={1 \over \sqrt {N}} \sum\nolimits\limits_{
j =0}^{N -1} {f}_{j}^{*}\left({{k}_{m}}\right)
{F}_{j} ,\quad m =0,1,\ldots , N -1
\label{2.11}
\end{equation}

The orthonormal set 
\begin{equation}
{\left\{{{1 \over \sqrt {N}} {f}_{j} \left({{
k}_{m}}\right)}\right\}}_{m =0}^{N -1}
\label{2.12}
\end{equation}
can be chosen as a basis for the Hilbert space $\not\subset\left(\left[0,
N\right] \right)$ , hence, a completness relation can be constructed
\begin{equation} {1 \over N\varepsilon} \sum\nolimits\limits_{m =0}^{ N -1}
{f}_{j}^{*}\left({{k}_{m}}\right) {f}_{j '}\left({{k}_{m}}\right) ={1 \over
\varepsilon} {\delta }_{jj'}
\label{2.13}
\end{equation}
the right hand side being the discrete analog of the Dirac Delta function.

Consider now the following function on the lattice
\begin{equation}
{\left.{{u}_{j} ={2 \over N\varepsilon i} \sum\nolimits\limits_{
m =1}^{s} {1 \over {k}_{m}} \left\{{{f}_{j}
\left({{k}_{m}}\right) -1}\right\}}\right|}_{s = N}
\label{2.14}
\end{equation}

Using the difference equations (\ref{2.2}) we can calculate $u_j$, namely,
\begin{eqnarray*}
 u_j &= &\left. {2 \over N\varepsilon i}
\sum\nolimits\limits_{m=1}^{s} {1 \over {k_m}}
\left\{{f_j-f_{j-1}+f_{j-1}-f_{j-2}+\ldots +f_2-f_1+f_1-f_0}\right\}
\right \vert_{s=N} = \\
& &\left. {2 \over N\varepsilon i}
\sum\nolimits\limits_{m=1}^{s} {ik_m\varepsilon \over 2k_m}
\left\{f_j+f_{j-1}+f_{j-1}+f_{j-2}+\ldots +f_2+f_1+f_1+f_0\right\}
\right \vert_{s=N} =1,\; \forall j>0
\end{eqnarray*}

Similarly, $u_j=-1,\quad \forall j<0$

Hence we can have the discrete analog of the Heaviside function,
\begin{equation}
\vartheta_j\equiv {1 \over 2}\left(u_j+1\right)=\left\{
\begin{array}{cll}
1&,&j>0\\
1/2&,&j=0\\
0&,&j<0
\end{array}\right.
\label{2.15}
\end{equation}

It is easy to prove the difference equations
\begin{eqnarray}
{1 \over \varepsilon} \Delta {\vartheta}_{j}&=&{1 \over
2 \varepsilon} \left({{\delta}_{j +1,0}+{\delta}_{j
,0}}\right),\qquad {1 \over \varepsilon} \nabla {\vartheta}_{j
}={1 \over 2 \varepsilon} \left({{\delta}_{j ,0}+{\delta
}_{j -1,0}}\right) \label{2.16}\\
{1 \over \varepsilon} \delta {\vartheta}_{j}&=&{1 \over
2 \varepsilon} \left({{\delta}_{j +{1 \over 2},0}+{\delta
}_{j -{1 \over 2},0}}\right)\label{2.17}
\end{eqnarray}
conecting the discrete versions of the Heaviside and Dirac functions.

We have also the useful relations:
\begin{eqnarray}
 \Delta \left({{f}_{j} {g}_{j}}\right) &=&
\left({\Delta{f}_{j}}\right)\left({\tilde{\Delta}{
g}_{j}}\right) +\left({\tilde{\Delta}{f}_{j}}\right) \left({\Delta
{g}_{j}}\right)
\label{2.18}\\
\tilde{\Delta}\left({{f}_{j} {g}_{j}}\right) &=&
\left({\tilde{\Delta}{f}_{j}}\right)\left({\tilde{\Delta}{
g}_{j}}\right) +{1 \over 4}\left({\Delta {f}_{j}}\right)
\left({\Delta {g}_{j}}\right) 
\label{2.19} 
\end{eqnarray}

\section{Hamiltonian formalism of the Klein-Gordon field on a
$(1+1)$-dimensional Minkowski lattice} 
We introduce the method of finite differences for the Klein-Gordon scalar
field. An explicit scheme for the wave equation consistent with the continuous
case (the truncation error is of second orden with respect to space and time
variables) can be constructed as follows:
\begin{equation}
\left({{1 \over {\tau}^{2}}{\nabla}_{n} {\Delta
}_{n} {\tilde{\nabla}}_{j} {\tilde{\Delta}}_{j} -{1 \over
{\varepsilon}^{2}}{\nabla}_{j} {\Delta}_{j}
{\tilde{\nabla}}_{n} {\tilde{\Delta}}_{n} +{M}^{
2}{\tilde{\nabla}}_{n} {\tilde{\Delta}}_{n} {\tilde{\nabla
}}_{j} {\tilde{\Delta}}_{j}}\right) {\phi}_{j}^{n}
=0
\label{3.1}
\end{equation}
where the field is defined in the grid points of the $(1+1)$-dimensional
lattice $ \phi_j^n \equiv \phi \left(j\varepsilon,n\tau\right),\;\varepsilon
, \tau $ being the space and time fundamental intervals, $j, n$ integer
numbers and ${\Delta}_{j} \left({{\nabla}_{j}}\right)$ are the
forward (backward) differences with respect to the space index, ${\tilde{
\Delta}}_{j} \left({{\tilde{\nabla}}_{j}}\right)$ the forward
(backward) averages, and similarly for the time index.

Using the method of separation of variables it can easily be proved that the
following functions of discrete variables are solutions of the wave equation
(\ref{3.1}):
\begin{equation}
{f}_{j}^{n} \left({k , \omega}\right)
={\left({{1+{1 \over 2} i \varepsilon k \over 1-{1
\over 2} i \varepsilon k}}\right)}^{j}
{\left({{1-{1 \over 2} i \tau \omega \over 1+{1 \over
2} i \tau \omega}}\right)}^{n}
\label{3.2}
\end{equation}
provided the ``dispersion relation'' is satisfied:
\begin{equation}
{\omega}^{2}-{k}^{2}={M}^{2}
\label{3.3}
\end{equation}
M, being the mass of the particle.

We have also the solutions of (\ref{3.1})
\begin{equation}
{\phi}_{j}^{n} ={\left({-1}\right)}^{j + n}
\label{3.4}
\end{equation}

In the limit, $j\rightarrow \infty,\quad n \rightarrow \infty,\quad 
j\varepsilon \rightarrow x,\quad n\tau \rightarrow t$ the functions (\ref{3.2})
become plane wave solutions
\begin{equation}
{f}_{j}^{n} \left({k , \omega}\right) \rightarrow
\exp i\left({kx - \omega t}\right)
\label{3.5}
\end{equation}

Imposing boundary conditions on the space indices
\begin{equation}
{f}_{o}^{n} \left({k , \omega}\right) ={f}_{
N}^{n} \left({k , \omega}\right)
\label{3.6}
\end{equation}
we get
\begin{equation}
k_m ={2 \over\varepsilon} \tan {\pi m \over N},\quad m=0,1,\ldots ,N-1
\label{3.7}
\end{equation}
therefore 
\begin{equation}
{\omega =\pm \left({{k}_{m}^{2}+{M}^{2}}\right)}^{1/2}
\label{3.8}
\end{equation}
For the positive energy solutions we define
\begin{equation}
{\omega_m = + \left({{k}_{m}^{2}+{M}^{2}}\right)}^{1/2}
\label{3.9}
\end{equation}

The solutions (\ref{3.4}) are a particular case of ${f}_{j}^{m}
\left({{k}_{m} ,{\omega}_{m}}\right)$ with $m=N/2$.

Starting from the wave equation (\ref{3.1}) we can construct a current vector.
Multiplying (\ref{3.1}) by ${\tilde{\nabla}}_{n} {\tilde{\Delta
}}_{n} {\tilde{\nabla}}_{j} {\tilde{\Delta}}_{j}
{\phi}_{j}^{* n}$ from the left, and multiplying the complex
conjugate of the wave equation by ${\tilde{\nabla}}_{n}
{\tilde{\Delta}}_{n} {\tilde{\nabla}}_{j} {\tilde{\Delta
}}_{j} {\phi}_{j}^{n}$ from the right, substracting both results
and using (\ref{2.18}) we obtain the ``conservation law''
\begin{equation}
{1 \over \varepsilon} {\nabla}_{j} {j}_{1}-
i {1 \over \tau} {\nabla}_{n} {j}_{4}=0
\label{3.10}
\end{equation}
where
\begin{eqnarray}
{j}_{1}&\equiv & i \left[{{1 \over \varepsilon} {\Delta
}_{j} \left({{\tilde{\nabla}}_{n} {\tilde{\Delta}}_{
n} {\phi}_{j}^{* n}}\right) {\tilde{\Delta}}_{
j} \left({{\tilde{\nabla}}_{n} {\tilde{\Delta}}_{n} {
\phi}_{j}^{n}}\right) -{\tilde{\Delta}}_{j} \left({{\tilde{\nabla
}}_{n} {\tilde{\Delta}}_{n} {\phi}_{j}^{*
n}}\right) {1 \over \varepsilon} {\Delta}_{j}
\left({{\tilde{\nabla}}_{n} {\tilde{\Delta}}_{n} {\phi
}_{j}^{n}}\right)}\right] 
\label{3.11}\\
{j}_{4}&\equiv & i \rho \equiv \left[{{1 \over \tau}
{\Delta}_{n} \left({{\tilde{\nabla}}_{j} {\tilde{\Delta
}}_{j} {\phi}_{j}^{* n}}\right) {\tilde{\Delta
}}_{n} \left({{\tilde{\nabla}}_{j} {\tilde{\Delta}}_{
j} {\phi}_{j}^{n}}\right) -{\tilde{\Delta}}_{n}
\left({{\tilde{\nabla}}_{j} {\tilde{\Delta}}_{j} {\phi
}_{j}^{* n}}\right) {1 \over \tau} {\Delta}_{n}
\left({{\tilde{\nabla}}_{j} {\tilde{\Delta}}_{j} {\phi
}_{j}^{n}}\right)}\right]
\label{3.12}
\end{eqnarray}
are the spatial and time component, respectively, of the charge vector current
on the lattice.

The charge density $\rho$ suggest that we can substitute the scalar field
$\phi \left({x , t}\right)$ by the promediated quantities ${\tilde{\nabla}}_{ j}
{\tilde{\Delta}}_{j} {\phi}_{j}^{n}$ and $\phi^*\left({x,t}\right)$ by
${\tilde{\nabla}}_{ j} {\tilde{\Delta}}_{j} {\phi}_{j}^{*n}$.

Now we address ourselves to the real Klein-Gordon field . A suitable
Hamiltonian for the field $\phi_j^n$ and its conjugate momentum $\pi_j^n$ can
be defined as follows
\begin{equation}
{H}_{n} = \varepsilon \sum\nolimits\limits_{j =0}^{N
-1} {1 \over 2}\left\{{{\left({{\tilde{\nabla}}_{j} {\tilde{\Delta
}}_{j} {\pi}_{j}^{n}}\right)}^{2}+{1 \over {\varepsilon
}^{2}}{\left({{\nabla}_{j} {\tilde{\nabla}}_{j} {
\phi}_{j}^{n}}\right)}^{2}+{M}^{2}{\left({{\tilde{\nabla}}_{
j} {\tilde{\Delta}}_{j} {\phi}_{j}^{n}}\right)}^{
2}}\right\}
\label{3.13}
\end{equation}
which is obviously hermitian. Using the periodicity condition of the fields and
the identity (\ref{2.18}) we can integrate by parts the second term on the right
with the result:
\begin{equation}
{H}_{n} = \varepsilon \sum\nolimits\limits_{j =0}^{N
-1} {1 \over 2}\left\{{{\left({{\tilde{\nabla}}_{j} {\tilde{\Delta
}}_{j} {\pi}_{j}^{n}}\right)}^{2}-{1 \over {\varepsilon
}^{2}}\left({{\nabla}_{j} {\Delta}_{j} {\phi
}_{j}^{n}}\right) \left({{\tilde{\nabla}}_{j} {\tilde{\Delta
}}_{j} {\phi}_{j}^{n}}\right) +{M}^{
2}{\left({{\tilde{\nabla}}_{j} {\tilde{\Delta}}_{j} {
\phi}_{j}^{n}}\right)}^{2}}\right\}\equiv \varepsilon 
\sum\nolimits\limits_{j =0}^{N -1} {\cal H}_{j}^{ n}
\label{3.14}
\end{equation}

As in the continuous case, we can derived the Hamilton equations of motions,
varying the Hamiltonian density ${\cal H}_j^n$ first with respecto to the
promediate momentum and secondly with respect to scalar field:
\begin{eqnarray}
{1 \over \tau} {\Delta}_{n} \left({{\tilde{\nabla
}}_{j} {\tilde{\Delta}}_{j} {\phi}_{j}^{n}}\right) &=&
{\partial {\cal H}_{j}^{ n} \over \partial \left({{\tilde{\nabla
}}_{j} {\tilde{\Delta}}_{j} {\pi}_{j}^{n}}\right)}
={\tilde{\Delta}}_{n} \left({{\tilde{\nabla}}_{j}
{\tilde{\Delta}}_{j} {\pi}_{j}^{n}}\right) 
\label{3.15} \\
{1 \over \tau} {\Delta}_{n} \left({{\tilde{\nabla
}}_{j} {\tilde{\Delta}}_{j} {\pi}_{j}^{n}}\right) &=&
{}-{\partial {\cal H}_{j}^{ n} \over \partial \left({{\tilde{\nabla
}}_{j} {\tilde{\Delta}}_{j} {\phi}_{j}^{n}}\right)}
={\tilde{\Delta}}_{n} \left({{1 \over {\varepsilon}^{
2}}{\nabla}_{j} {\Delta}_{j} {\phi}_{j}^{n} -{
M}^{2}{\tilde{\nabla}}_{j} {\tilde{\Delta}}_{j} {
\phi}_{j}^{n}}\right)
\label{3.16}
\end{eqnarray}

(The same result was obtained by Bender and Milton (1986) if one eliminates the
auxiliary field $\Gamma$ in their formule (2.5) of Reference 8).

Applying the difference operator ${1 \over \tau} {\nabla}_{
n}$ on both sides of (\ref{3.15}) and substituying (\ref{3.16}) in the result we
recover the wave equation (\ref{3.1}).

Using (\ref{3.15}) and (\ref{3.16}) it can easily be proved that the Hamiltonian
(\ref{3.14}) is independent of the time index $n$, namely:
\begin{equation}
{\nabla}_{n} {H}_{n} ={\Delta}_{n} {
H}_{n} =0
\label{3.17}
\end{equation}
hence
\begin{equation}
{\tilde{\nabla}}_{n} {H}_{n} ={\tilde{\Delta}}_{
n} {H}_{n} ={H}_{n}
\label{3.18}
\end{equation}

Since the plane wave solutions ${f}_{j}^{n} \left({{k}_{m}
,{\omega}_{m}}\right) \left({m =0,1\ldots N
-1}\right)$ form a complete set of orthogonal functions, we
can expand the wave field and its conjugate momentum as
\begin{eqnarray}
{\phi}_{j}^{n} &=&{1 \over \sqrt {N\varepsilon}}
\sum\nolimits\limits_{m =- N/ 2}^{N/ 2-1} {\left({1+{1
\over 4}{\varepsilon}^{2}{k}_{m}^{2}}\right) \over \sqrt
{2{\omega}_{m}}} \left({{a}_{m} {f}_{j}^{n}
\left({{k}_{m} ,{\omega}_{m}}\right) +{a}_{m}^{
*}{f}_{j}^{* n} \left({{k}_{m} ,{\omega
}_{m}}\right)}\right)
\label{3.19} \\
{\pi}_{j}^{n} &=&{- i \over \sqrt {N\varepsilon}}
\sum\nolimits\limits_{m =- N/ 2}^{N/ 2-1} \sqrt {{{
\omega}_{m} \over 2}}\left({1+{1 \over 4}{\varepsilon}^{2}{
k}_{m}^{2}}\right)\left({{a}_{m} {f}_{j}^{n} \left({{
k}_{m} ,{\omega}_{m}}\right) -{a}_{m}^{*}{f}_{
j}^{* n} \left({{k}_{m} ,{\omega
}_{m}}\right)}\right)
\label{3.20}
\end{eqnarray}

Applying the average operator on both sides we get
\begin{eqnarray}
{\tilde{\nabla}}_{j} {\tilde{\Delta}}_{j} {\phi
}_{j}^{n} &=&{1 \over \sqrt {N\varepsilon}}
\sum\nolimits\limits_{m =- N/ 2}^{N/ 2-1} {1 \over \sqrt
{2{\omega}_{m}}} \left({{a}_{m} {f}_{j}^{n}
\left({{k}_{m} ,{\omega}_{m}}\right) +{a}_{m}^{
*}{f}_{j}^{* n} \left({{k}_{m} ,{\omega
}_{m}}\right)}\right)
\label{3.21} \\
{\tilde{\nabla}}_{j} {\tilde{\Delta}}_{j} {\pi
}_{j}^{n} &=&{- i \over \sqrt {N\varepsilon}}
\sum\nolimits\limits_{m =- N/ 2}^{N/ 2-1} \sqrt {{{
\omega}_{m} \over 2}}\left({{a}_{m} {f}_{j}^{n}
\left({{k}_{m} ,{\omega}_{m}}\right) -{a}_{m}^{
*}{f}_{j}^{* n} \left({{k}_{m} ,{\omega
}_{m}}\right)}\right)
\label{3.22}
\end{eqnarray}

Using the completness relation (\ref{2.13}) we can calculate the coefficients in
the Fourier expansion
\begin{eqnarray}
{a}_{m} &=&{\varepsilon \over \sqrt {N\varepsilon}}
\sum\nolimits\limits_{j =0}^{N -1} \left\{{\sqrt {{{\omega
}_{m} \over 2}}{f}_{j}^{* n} \left({{k}_{m}
,{\omega}_{m}}\right) {\tilde{\nabla}}_{j} {\tilde{\Delta
}}_{j} {\phi}_{j}^{n} +{i \over \sqrt {2{\omega
}_{m}}} {f}_{j}^{* n} \left({{k}_{m} ,{
\omega}_{m}}\right) {\tilde{\nabla}}_{j} {\tilde{\Delta}}_{
j} {\pi}_{j}^{n}}\right\}
\label{3.23} \\
{a}_{m}^{*}&=&{\varepsilon \over \sqrt {N\varepsilon}}
\sum\nolimits\limits_{j =0}^{N -1} \left\{{\sqrt {{{\omega
}_{m} \over 2}}{f}_{j}^{n} \left({{k}_{m} ,{\omega
}_{m}}\right) {\tilde{\nabla}}_{j} {\tilde{\Delta}}_{j}
{\phi}_{j}^{n} -{i \over \sqrt {2{\omega}_{m}}}
{f}_{j}^{n} \left({{k}_{m} ,{\omega}_{m}}\right)
{\tilde{\nabla}}_{j} {\tilde{\Delta}}_{j} {\pi
}_{j}^{n}}\right\}
\label{3.24}
\end{eqnarray}

Using (\ref{2.18}), the periodicity condition and Hamilton equations of motions
(\ref{3.15}, \ref{3.16}), one can easily prove that this coefficients are
independent of $n$, namely,
\begin{equation}
{\nabla}_{n} \left({{a}_{m}}\right) ={\Delta}_{n}
\left({{a}_{m}}\right) ={\nabla}_{n} \left({{a}_{m}^{
*}}\right)={\Delta}_{n} \left({{a}_{m}^{*}}\right)=0
\label{3.25}
\end{equation}

Turning to the complex Klein-Gordon field we can define a Hamiltonian as
follows:
\begin{equation}
{H}_{n} = \varepsilon \sum\nolimits\limits_{j =0}^{N
-1} \left\{{\left({{\tilde{\nabla}}_{j} {\tilde{\Delta}}_{j}
{\pi}^{*}}\right)\left({{\tilde{\nabla}}_{j}
{\tilde{\Delta}}_{j} {\pi}_{j}^{n}}\right)
-\left({{\tilde{\nabla}}_{j} {\tilde{\Delta}}_{j} {\phi
}_{j}^{* n}}\right) {1 \over {\varepsilon}^{2}}{\nabla
}_{j} {\Delta}_{j} {\phi}_{j}^{n} +{M}^{
2}\left({{\tilde{\nabla}}_{j} {\tilde{\Delta}}_{j} {\phi
}_{j}^{* n}}\right) {\tilde{\nabla}}_{j} {\tilde{\Delta
}}_{j} {\phi}_{j}^{n}}\right\}
\label{3.26}
\end{equation}
from which the Hamilton equations of motion can be derived:
\begin{eqnarray}
{1 \over \tau} {\Delta}_{n} {\tilde{\nabla}}_{j}
{\tilde{\Delta}}_{j} {\phi}_{j}^{n} &=&{\tilde{\Delta}}_{
n} \left({{\tilde{\nabla}}_{j} {\tilde{\Delta}}_{j} {
\pi}_{j}^{* n}}\right) 
\label{3.27}\\
{1 \over \tau} {\Delta}_{n} \left({{\tilde{\nabla
}}_{j} {\tilde{\Delta}}_{j} {\pi}_{j}^{n}}\right)
&=&{\tilde{\Delta}}_{n} \left({{1 \over {\varepsilon}^{
2}}{\nabla}_{j} {\Delta}_{j} {\phi}_{j}^{*
n} -{M}^{2}\left({{\tilde{\nabla}}_{j} {\tilde{\Delta
}}_{j} {\phi}_{j}^{* n}}\right)}\right) 
\label{3.28}
\end{eqnarray}
leading to the wave equation (\ref{3.1}). The solution of the wave equation can
be expanded in Fourier series as before:
\begin{eqnarray*}
{\phi}_{j}^{n} &=&{1 \over \sqrt {N\varepsilon}}
\sum\nolimits\limits_{m =- N/ 2}^{N/ 2-1} {\left({1+{1
\over 4}{\varepsilon}^{2}{k}_{m}^{2}}\right) \over \sqrt
{2{\omega}_{m}}} \left({{a}_{m} {f}_{j}^{n}
\left({{k}_{m} ,{\omega}_{m}}\right) +{b}_{m}^{
*}{f}_{j}^{* n} \left({{k}_{m} ,{\omega
}_{m}}\right)}\right) \\
{\pi}_{j}^{n} &=&{- i \over \sqrt {N\varepsilon
}} \sum\nolimits\limits_{m =- N/ 2}^{N/ 2-1} \sqrt
{{{\omega}_{m} \over 2}}\left({1+{1 \over 4}{\varepsilon}^{
2}{k}_{m}^{2}}\right)\left({{b}_{m} {f}_{j}^{n}
\left({{k}_{m} ,{\omega}_{m}}\right) -{a}_{m}^{*}{f}_{j}^{*n} \left({{k}_{m}
,{\omega}_{m}}\right)}\right) \end{eqnarray*}
and similar expressions for the complex fields.

In order to make connection of our scheme with the Einstein-de Broglie
relations $E=\hbar \omega, \quad p=\hbar k$ we take for the period $T$ and
wave length $\lambda$ of the discrete plane waves functions (\ref{3.2}) and
(\ref{3.6})
$$T=N\tau, \qquad \lambda = N\varepsilon $$
and for the phase velocity
$${v}_{p} ={\lambda \over T} ={\varepsilon \over \tau}$$

We have define the wave number and the angular frequency of the wave functions
as:
$${k}_{m} ={2 \over \varepsilon} \tan{\pi m \over N},
\quad {\omega}_{m} ={2 \over \tau} \tan{\pi m \over N}
,\quad m =0,1,\ldots , N -1$$
substituting the Einstein-de Broglie relations in the relativistic expresion 
$E^2-p^2=M^2$ (we use natural units $\hbar=c=1$), we obtain
$${\omega}_{m}^{2}-{k}_{m}^{2}={\omega}_{m}^{
2}\left({1-{{\tau}^{2} \over {\varepsilon}^{
2}}}\right)={\omega}_{m}^{2}\left({1-{1 \over {v}_{p}^{
2}}}\right)={M}^{2}$$

Since the phase velocity and group velocity satisfy $v_pv_g=1$, we have finally
$${\omega}_{m}^{2}={{M}^{2} \over 1-{v}_{g}^{2}}$$

\section{Quantitation of the Klein-Gordon field}
Let us start form the Hamilton equation of motion (\ref{3.15}, \ref{3.16}) where
the real fields $\phi_n^n$ and $\pi_n^n$ now are, in general, non commuting
operators:
\begin{eqnarray}
{1 \over \tau} {\Delta}_{n} \left({{\tilde{\nabla
}}_{j} {\tilde{\Delta}}_{j} {\phi}_{j}^{n}}\right)
&=&{\tilde{\Delta}}_{n} \left({{\tilde{\nabla}}_{j}
{\tilde{\Delta}}_{j} {\pi}_{j}^{n}}\right) 
\label{4.1} \\
{1 \over \tau} {\Delta}_{n} \left({{\tilde{\nabla
}}_{j} {\tilde{\Delta}}_{j} {\pi}_{j}^{n}}\right)
&=&{\tilde{\Delta}}_{n} \left({{1 \over {\varepsilon}^{
2}}{\nabla}_{j} {\Delta}_{j} {\phi}_{j}^{n} -{
M}^{2}\left({{\tilde{\nabla}}_{j} {\tilde{\Delta}}_{j}
{\phi}_{j}^{n}}\right)}\right)
\label{4.2}
\end{eqnarray}

Eliminating the field $\pi_j^n$ we obtain again the wave equation on the lattice
(\ref{3.1}) the solutions of which can be expanded in terms of the ``plane wave
functions''
\begin{eqnarray}
{\tilde{\nabla}}_{j} {\tilde{\Delta}}_{j} {\phi
}_{j}^{n} &=&{1 \over \sqrt {N\varepsilon}}
\sum\nolimits\limits_{m =- N/ 2}^{N/ 2-1} {1 \over \sqrt
{2{\omega}_{m}}}\left({{a}_{m} {f}_{j}^{n} \left({{
k}_{m} ,{\omega}_{m}}\right) +{a}_{m}^{\dag}{f}_{
j}^{* n} \left({{k}_{m} ,{\omega
}_{m}}\right)}\right)
\label{4.3} \\
{\tilde{\nabla}}_{j} {\tilde{\Delta}}_{j} {\pi
}_{j}^{n} &=&{- i \over \sqrt {N\varepsilon}}
\sum\nolimits\limits_{m =- N/ 2}^{N/ 2-1} \sqrt {{{
\omega}_{m} \over 2}}\left({{a}_{m} {f}_{j}^{n}
\left({{k}_{m} ,{\omega}_{m}}\right) -{a}_{m}^{\dag
}{f}_{j}^{* n} \left({{k}_{m} ,{\omega
}_{m}}\right)}\right)
\label{4.4}
\end{eqnarray}
where the coefficients $a_m, \quad a_m^\dagger$, are, in general,
non-commuting operators.

We introduce the quantization procedure imposing the commutation relations of
the fields for $n = 0$
\begin{eqnarray}
\left[{{\tilde{\nabla}}_{j} {\tilde{\Delta}}_{j} {
\phi}_{j}^{0},{\tilde{\nabla}}_{j'}{\tilde{\Delta}}_{j
'}{\phi}_{j'}^{0}}\right]&=&0
\label{4.5} \\
\left[{{\tilde{\nabla}}_{j} {\tilde{\Delta}}_{j} {
\pi}_{j}^{0},{\tilde{\nabla}}_{j'}{\tilde{\Delta}}_{j
'}{\pi}_{j'}^{0}}\right]&=&0
\label{4.6} \\
 \left[{{\tilde{\nabla}}_{j} {\tilde{\Delta}}_{j} {
\phi}_{j}^{0},{\tilde{\nabla}}_{j'}{\tilde{\Delta}}_{j
'}{\pi}_{j'}^{0}}\right]&=&{i \over \varepsilon} {\delta
}_{jj'}
\label{4.7}
\end{eqnarray}

In order to prove unitarity we must prove that these relations hold also for
any $n$. Since (\ref{4.3}) and (\ref{4.4}) are valid for $n = 0$, we solve them
for $a_m$ and $a_m^\dagger$
\begin{eqnarray}
{\left({{a}_{m}}\right)}^{0}&=&{\varepsilon \over \sqrt {N\varepsilon
}} \sum\nolimits\limits_{j =0}^{N -1} \left\{{\sqrt {{{
\omega}_{m} \over 2}}{f}_{j}^{*0}\left({{k}_{m} ,{
\omega}_{m}}\right) {\tilde{\nabla}}_{j} {\tilde{\Delta}}_{
j} {\phi}_{j}^{0}+{i \over \sqrt {2{\omega
}_{m}}} {f}_{j}^{*0}\left({{k}_{m} ,{\omega
}_{m}}\right) {\tilde{\nabla}}_{j} {\tilde{\Delta}}_{j}
{\pi}_{j}^{0}}\right\}
\label{4.8} \\
{\left({{a}_{m}^{\dag}}\right)}^{0}&=&{\varepsilon \over \sqrt
{N\varepsilon}} \sum\nolimits\limits_{j =0}^{N -1}
\left\{{\sqrt {{{\omega}_{m} \over 2}}{f}_{j}^{
0}\left({{k}_{m} ,{\omega}_{m}}\right) {\tilde{\nabla}}_{
j} {\tilde{\Delta}}_{j} {\phi}_{j}^{0}-{i \over
\sqrt {2{\omega}_{m}}} {f}_{j}^{0}\left({{k}_{m}
,{\omega}_{m}}\right) {\tilde{\nabla}}_{j} {\tilde{\Delta
}}_{j} {\pi}_{j}^{0}}\right\}
\label{4.9}
\end{eqnarray}

Using (\ref{4.5}--\ref{4.7}) it is straight forward to prove that
\begin{eqnarray}
\biggl[{\Bigl( a_m \Bigr)}^0,{\Bigl( a_{m'} \Bigr)}^0\biggr] &= &0,\qquad
\biggl[{\Bigl( a_m^{\dag} \Bigr)}^0 , {\Bigl( a_{m'}^{\dag} \Bigr)}^0\biggr]=0
\label{4.10} \\ 
\biggl[{\Bigl( a_m \Bigr)}^0, {\Bigl(
a_{m'}^{\dag}\Bigr)}^0\biggr]&=&\delta_{mm'} \label{4.11} \end{eqnarray}

From (\ref{4.1}, \ref{4.2}) and (\ref{4.8}, \ref{4.9}) one can prove (remember
\ref{3.25})
\begin{equation}
{\Bigl( a_m \Bigr)}^0={\Bigl( a_m \Bigr)}^1,\quad
{\Bigl( a_m^{\dag} \Bigr)}^0={\Bigl( a_m^{\dag} \Bigr)}^1
\label{4.12}
\end{equation}
hence (\ref{4.10}, \ref{4.11}) is also true for $n = 1$, and from (\ref{4.3},
\ref{4.4}) we derive
\begin{eqnarray}
\left[{{\tilde{\nabla}}_{j} {\tilde{\Delta}}_{j} {
\phi}_{j}^{1},{\tilde{\nabla}}_{j'}{\tilde{\Delta}}_{j
'}{\phi}_{j'}^{1}}\right]&=&\left[{{\tilde{\nabla}}_{j}
{\tilde{\Delta}}_{j} {\pi}_{j}^{1},{\tilde{\nabla}}_{
j'}{\tilde{\Delta}}_{j'}{\pi}_{j'}^{1}}\right]=0
\label{4.13} \\
\left[{{\tilde{\nabla}}_{j} {\tilde{\Delta}}_{j} {
\phi}_{j}^{1},{\tilde{\nabla}}_{j'}{\tilde{\Delta}}_{j
'}{\pi}_{j'}^{1}}\right]&=&{i \over \varepsilon} {\delta
}_{jj'}
\label{4.14}
\end{eqnarray}
and the proof is completed by induction.

The Hamiltonian for the real scalar field reads:
\begin{equation}
{H}_{n} = \varepsilon \sum\nolimits\limits_{j =0}^{N
-1} {1 \over 2}\left\{{{\left({{\tilde{\nabla}}_{j} {\tilde{\Delta
}}_{j} {\pi}_{j}^{n}}\right)}^{2}-{1 \over {\varepsilon
}^{2}}\left({{\nabla}_{j} {\Delta}_{j} {\phi
}_{j}^{n}}\right) \left({{\tilde{\nabla}}_{j} {\tilde{\Delta
}}_{j} {\phi}_{j}^{n}}\right) +{M}^{
2}{\left({{\tilde{\nabla}}_{j} {\tilde{\Delta}}_{j} {
\phi}_{j}^{n}}\right)}^{2}}\right\}
\label{4.15}
\end{equation}

A convenient scheme for the Heisenberg equation is:
\begin{eqnarray*}
{1 \over \tau} {\Delta}_{n} \left({{\tilde{\nabla
}}_{j} {\tilde{\Delta}}_{j} {\phi}_{j}^{n}}\right)
&=&{1 \over i} \left[{{\tilde{\Delta}}_{n}
\left({{\tilde{\nabla}}_{j} {\tilde{\Delta}}_{j} {\phi
}_{j}^{n}}\right) ,{H}_{n}}\right] \\
{1 \over \tau} {\Delta}_{n} \left({{\tilde{\nabla
}}_{j} {\tilde{\Delta}}_{j} {\pi}_{j}^{n}}\right)
&=&{1 \over i} \left[{{\tilde{\Delta}}_{n}
\left({{\tilde{\nabla}}_{j} {\tilde{\Delta}}_{j} {\pi
}_{j}^{n}}\right) ,{H}_{n}}\right]
\end{eqnarray*}

Using (\ref{2.19}) and (\ref{3.18}) and (\ref{4.1}, \ref{4.2}) for any $n$, we
get for the first equation
\begin{equation}
{1 \over \tau} {\Delta}_{n} \left({{\tilde{\nabla
}}_{j} {\tilde{\Delta}}_{j} {\phi}_{j}^{n}}\right)
={1 \over i} {\tilde{\Delta}}_{n} \left[{{\tilde{\nabla
}}_{j} {\tilde{\Delta}}_{j} {\phi}_{j}^{n} ,{
H}_{n}}\right] ={\tilde{\Delta}}_{n} \left({{\tilde{\nabla}}_{
j} {\tilde{\Delta}}_{j} {\pi}_{j}^{n}}\right) 
\label{4.16}
\end{equation}
and similarly for the second equation
\begin{equation}
{1 \over \tau} {\Delta}_{n} \left({{\tilde{\nabla
}}_{j} {\tilde{\Delta}}_{j} {\pi}_{j}^{n}}\right)
={\tilde{\Delta}}_n \left\{{{1 \over {\varepsilon}^{2}}{\nabla
}_{j} {\Delta}_{j} {\phi}_{j}^{n} -{M}^{
2}{\tilde{\nabla}}_{j} {\tilde{\Delta}}_{j} {\phi
}_{j}^{n}}\right\} 
\label{4.17}
\end{equation}
consistent with the Hamilton equation of motions (\ref{4.1}, \ref{4.2}).

Substituting (\ref{4.8}, \ref{4.9}) in (\ref{4.15}) we find
\begin{equation}
H ={1 \over 2}\sum\nolimits\limits_{m =- N/ 2}^{N/
2-1} {\omega}_{m} \left({{a}_{m} {a}_{m}^{\dag
}+{a}_{m}^{\dag}{a}_{m}}\right)
\label{4.18}
\end{equation}

If we define the linear momentum on the lattice as
$$P = \varepsilon \sum\nolimits\limits_{j =0}^{N -1}
{1 \over i\varepsilon} \left({{\tilde{\nabla}}_{j}
{\tilde{\Delta}}_{j} {\pi}_{j}^{n}}\right)
\left({{\tilde{\nabla}}_{j} {\Delta}_{j} {\phi
}_{j}^{n}}\right)$$
we find 
$$P =\sum\nolimits\limits_{m =- N/ 2}^{N/ 2-1} {
k}_{m} \left({{a}_{m} {a}_{m}^{\dag}+{a}_{m}^{
\dag}{a}_{m}}\right)=\sum\nolimits\limits_{m =- N/
2}^{N/ 2-1} {k}_{m} {a}_{m}^{\dag}{a}_{m} $$
because the zero point momentum vanish by cancellation of $k_m$ with $k_{-m}$ 

For the complex Klein-Gordon field we have the hermitian Hamiltonian
\begin{equation}
{H}_{n} = \varepsilon \sum\nolimits\limits_{j =0}^{N
-1} \left\{{{\left({{\tilde{\nabla}}_{j} {\tilde{\Delta}}_{j}
{\pi}^{\dag}}\right)\left({{\tilde{\nabla}}_{j}
{\tilde{\Delta}}_{j} {\pi}_{j}^{n}}\right)}
-\left({{\tilde{\nabla}}_{j} {\tilde{\Delta}}_{j} {\phi
}_{j}^{\dag n}}\right) {1 \over {\varepsilon}^{
2}}\left({{\nabla}_{j} {\Delta}_{j} {\phi
}_{j}^{n}}\right) +{M}^{2}{\left({{\tilde{\nabla}}_{j}
{\tilde{\Delta}}_{j} {\phi}_{j}^{\dag n}}\right)
\left({{\tilde{\nabla}}_{j} {\tilde{\Delta}}_{j} {\phi
}_{j}^{n}}\right)}}\right\} 
\label{4.20}
\end{equation}

The equal time commutation relations now read:
\begin{equation}
\left[{{\tilde{\nabla}}_{j} {\tilde{\Delta}}_{j} {
\phi}_{j}^{n} ,{\tilde{\nabla}}_{j'}{\tilde{\Delta}}_{j
'}{\pi}_{j'}^{n}}\right] ={1 \over \varepsilon}
{\delta}_{jj'}\quad ,\quad \left[{{\tilde{\nabla}}_{j}
{\tilde{\Delta}}_{j} {\phi}_{j}^{\dag n}
,{\tilde{\nabla}}_{j'}{\tilde{\Delta}}_{j'}{\pi
}_{j'}^{\dag n}}\right] ={1 \over \varepsilon} {
\delta}_{jj'} \label{4.21}
\end{equation}
with all other equal time commutators vanishing.

The Heisenberg equations of motion lead, as in the real case, to the discrete
wave equation for the fields $\phi_j^n, \; \pi_j^n$ , and their adjoints. The
Fourier expansion in terms of the plane wave solutions are
\begin{eqnarray}
{\phi}_{j}^{n} &=&{1 \over \sqrt {N\varepsilon}}
\sum\nolimits\limits_{m =- N/ 2}^{N/ 2-1} {\left({1+{1
\over 4}{\varepsilon}^{2}{k}_{m}}\right) \over \sqrt {{2
\omega}_{m}}} \left({{a}_{m} {f}_{j}^{n} \left({{
k}_{m} ,{\omega}_{m}}\right) +{b}_{m}^{\dag}{f}_{
j}^{* n} \left({{k}_{m} ,{\omega
}_{m}}\right)}\right)
\label{4.22} \\
{\pi}_{j}^{n} &=&{- i \over \sqrt {N\varepsilon}}
\sum\nolimits\limits_{m =- N/ 2}^{N/ 2-1} \sqrt {{{
\omega}_{m} \over 2}}\left({1+{1 \over 4}{\varepsilon}^{2}{
k}_{m}}\right) \left({{b}_{m} {f}_{j}^{n} \left({{
k}_{m} ,{\omega}_{m}}\right) -{a}_{m}^{\dag}{f}_{
j}^{* n} \left({{k}_{m} ,{\omega
}_{m}}\right)}\right)
\label{4.23}
\end{eqnarray}

Inverting this equations we can calculate with the help of (\ref{4.21}).
\begin{equation}
\left[{{a}_{m} ,{a}_{m'}^{\dag}}\right]={\delta}_{mm
'}\quad , \quad \left[{{b}_{m} ,{b}_{m'}^{\dag}}\right]={\delta
}_{mm'}
\label{4.24}
\end{equation}
with all other commutators vanishing.

The Hamiltonian can be expressed in terms of these creation and annihilation
operators:
\begin{equation}
H ={1 \over 2}\sum\nolimits\limits_{m =- N/ 2}^{N/ 2-1}{\omega}_{m} \left({{
a}_{m}^{\dag}{a}_{m} +{b}_{m}^{\dag}{
b}_{m}}\right)
\label{4.25}
\end{equation}
The charge of the complex field is given by
\begin{equation}
Q =- i\varepsilon \sum\nolimits\limits_{j =0}^{N -1}
\left({{\tilde{\nabla}}_{j} {\tilde{\Delta}}_{j} {\pi
}_{j}^{n}}\right) \left({{\tilde{\nabla}}_{j} {\tilde{\Delta
}}_{j} {\phi}_{j}^{n}}\right) -\left({{\tilde{\nabla}}_{
j} {\tilde{\Delta}}_{j} {\pi}_{j}^{\dag 
n}}\right) \left({{\tilde{\nabla}}_{j} {\tilde{\Delta}}_{
j} {\phi}_{j}^{\dag n}}\right) =\sum\nolimits\limits_{
m =- N/ 2}^{N/ 2-1} \left({{a}_{m}^{\dag}{
a}_{m} +{b}_{m}^{\dag}{b}_{m}}\right)
\label{4.26}
\end{equation}

Finally, two very important quantities can be evaluated on the lattice: the
general commutations rules and the Feynman propagator.

First of all, using (\ref{4.22}) and (\ref{4.24}) in order to calculate the
commutator of the fields for two different space-time points, we find:
\begin{equation}
\left[{{\tilde{\nabla}}_{j} {\tilde{\Delta}}_{j} {
\phi}_{j}^{n} ,{\tilde{\nabla}}_{j'}{\tilde{\Delta}}_{j
'}{\phi}_{j'}^{\dag n'}}\right]={1 \over N\varepsilon
} \sum\nolimits\limits_{m =- N/ 2}^{N/ 2-1} {1 \over
2{\omega}_{m}} \left({{f}_{j - j'}^{n -
n'}\left({{k}_{m} ,{\omega}_{m}}\right) -{f}_{j
- j'}^{* n - n'}\left({{k}_{m} ,{\omega
}_{m}}\right)}\right)
\label{4.27}
\end{equation}
which becomes zero for equal time $(n = n')$

Secondly, the Feynman propagator function is defined to be the vacuum
expectation value
\begin{equation}
{\Delta}_{F} \equiv \left\langle 0 \left\vert T {\phi
}_{j}^{n} {\phi}_{j'}^{n'}\right\vert 0 \right\rangle 
\label{4.28}
\end{equation}
of the time ordered product
\begin{equation}
T\phi_j^n \phi_{j'}^{\dag n'}=\left\{
\begin{array}{ll}
\phi_j^n\phi_{j'}^{\dag n'}&\mbox{for} \quad n>n'\\ 
\noalign{\medskip}
\phi_{j'}^{\dag n'}\phi_j^n &\mbox{for} \quad n'> n
\end{array}\right.
\label{4.29}
\end{equation}

Notice that there is no ambiguity of the time ordered product for equal times,
sine $\phi_j^n$ and $\phi_j^{\dag n'}$ commute for $n = n'$ because of
(\ref{4.27}).

The discrete analog of the Feynman propagator is from (\ref{4.22}) and
(\ref{4.29})
\begin{equation}
{\Delta}_{F} ={1 \over N\varepsilon}
\sum\nolimits\limits_{m =- N/ 2}^{N/ 2-1} {{\left({1+{1
\over 4}{\varepsilon}^{2}{k}_{m}}\right)}^{2} \over 2{
\omega}_{m}} \left\{{{\vartheta}_{n} {f}_{j}^{n}
\left({{k}_{m} ,{\omega}_{m}}\right) +{\vartheta}_{
- n} {f}_{j}^{*}\left({{k}_{m} ,{\omega
}_{m}}\right)}\right\}
\label{4.30}
\end{equation}
$\vartheta_n$ being the discrete realization of the Heaviside functions
(\ref{2.15}).

In order to see the analogy with the continuous case we could take the discrete
Fourier transform of (\ref{4.30}). Instead, we take the Fourier transform of the
wave equation (\ref{3.1}), namely,
$$\left({-{\omega}_{r}^{2}+{k}_{m}^{2}+{M}^{
2}}\right){1 \over 1+{1 \over 4}{\varepsilon}^{2}{k}_{m}^{
2}}{1 \over 1+{1 \over 4}{\tau}^{2}{\omega}_{m}^{
2}}{\hat{\phi}}_{m}^{r} =0$$
where 
$${\hat{\phi}}_{m}^{r} =\sum\nolimits\limits_{j =0}^{N -1}
\sum\nolimits\limits_{n =0}^{N -1} {f}_{j}^{*
n} \left({{k}_{m} ,{\omega}_{r}}\right) {\phi
}_{j}^{n}$$
is the Fourier transform of the Klein-Gordon field.

Therefore the Feynman propagator is defined in the momentum space
$${\hat{G}}_{m}^{r} =- i {\left({1+{1 \over 4}{\varepsilon
}^{2}{k}_{m}^{2}}\right)\left({1+{1 \over 4}{\tau}^{
2}{\omega}_{r}^{2}}\right) \over {k}_{m}^{2}-{\omega
}_{r}^{2}+{M}^{2}}$$
and in the configuration space
$${G}_{j - j'}^{n - n'}={1 \over \left({
N\varepsilon}\right) \left({N\tau}\right)}
\sum\nolimits\limits_{m =- N/ 2}^{N/ 2-1}
\sum\nolimits\limits_{r =- N/ 2}^{N/ 2-1} {\hat{
G}}_{m}^{r} {f}_{j - j'}^{n - n
'}\left({{k}_{m} ,{\omega}_{r}}\right) $$
(Notice that now ${\omega}_{r} ={2 \over \tau} \tan{\pi 
 r \over N}$, can take negative values)

It can be proved that this propagator is the Green fuction for the discrete
wave equation, namely,
$$\left({{{1 \over {\tau}^{2}}{\nabla}_{n} {\Delta
}_{n} {\tilde{\nabla}}_{j} {\tilde{\Delta}}_{j} -{1
\over {\varepsilon}^{2}}{\nabla}_{j} {\Delta}_{j}
{\tilde{\nabla}}_{n} {\tilde{\Delta}}_{n}}+{M}^{
2}{{\tilde{\nabla}}_{n} {\tilde{\Delta}}_{n} {\tilde{\nabla
}}_{j} {\tilde{\Delta}}_{j}}}\right){G}_{j -
j'}^{n - n'}={i \over \tau} {\delta}_{nn
'}{1 \over \varepsilon} {\delta}_{jj'}$$

With the aid of the identities:
\begin{eqnarray*}
{\nabla}_{n} {\Delta}_{n} \left[{{\vartheta}_{n}
{f}_{j}^{n}}\right] &=&{\vartheta}_{n} {\nabla}_{n}
{\Delta}_{n} {f}_{j}^{n} +\left({{\Delta}_{n} {
\vartheta}_{n}}\right) \left({{\Delta}_{n} {f}_{
j}^{n}}\right) +\left({{\Delta}_{n} {\vartheta}_{n
-1}}\right)\left({{\Delta}_{n} {f}_{j}^{n
-1}}\right)+\left({{\nabla}_{n} {\Delta}_{n} {\vartheta
}_{n}}\right) {f}_{j}^{n}
\\
{\tilde{\nabla}}_{n} {\tilde{\Delta}}_{n} \left[{{
\vartheta}_{n} {f}_{j}^{n}}\right] &=&{\vartheta}_{n}
{\tilde{\nabla}}_{n} {\tilde{\Delta}}_{n} {f}_{j}^{n}
+{1 \over 4}\left[{\left.{\left({{\Delta}_{n} {\vartheta
}_{n}}\right) \left({{\Delta}_{n} {f}_{j}^{n}}\right)
+\left({{\Delta}_{n} {\vartheta}_{n
-1}}\right)}\right.\left({{\Delta}_{n} {f}_{j}^{n
-1}}\right)}\right]+{1 \over 4}\left({{\nabla}_{n} {\Delta}_{
n} {\vartheta}_{n}}\right) {f}_{j}^{n}
\end{eqnarray*}
and the equations (\ref{2.16}) and (\ref{3.3}) one can prove that
$\Delta_F$ defined by (\ref{4.30}) satisfies also the same condition,
therefore $\Delta_F$ and $G_j^n$ are equal up to a constant.

\section{Quantization of the Dirac field}
The Hamiltonian for the Dirac field ${\psi}_{\alpha j}^n$ and its hermitian
adjoint ${\psi}_{\alpha j}^{\dag n}$ can be defined
as:
\begin{equation}
{H}_{n}=\varepsilon \sum\nolimits\limits_{j=0}^{N
-1} {{\tilde{\Delta}}_{j}{\psi}_{j}^{\dag n}
\left\{{{\gamma}_{4}{\gamma}_{1}{1 \over {\varepsilon
}}{\Delta}_{j}{\psi}_{j}^{n}+M{\gamma
}_{4}{\tilde{\Delta}}_{j}{\psi}_{j}^{n}}\right\}}
\label{5.1}
\end{equation}
with 	
\begin{equation}
{\gamma}_{1}=\left({\matrix{0&-i\cr
i&0\cr}}\right),\quad {\gamma}_{4}=\left({\matrix{1&0\cr
0&-1\cr}}\right),\quad i{\gamma}_{1}{\gamma}_{4}={\gamma}_{5}=\left({\matrix{0&-1\cr
-1&0\cr}}\right)
\label{5.2}
\end{equation}

In the next chapter it will be shown that the substitution $\psi \left({\mit
x}\right)\rightarrow {\tilde{\Delta}}_{j}{\psi}_{j}^{n}$
is consistent with the density function
\begin{equation}
\rho =\left({{\tilde{\Delta}}_{j}{\psi}_{j}^{\dag \mit
n}}\right)\left({{\tilde{\Delta}}_{j}{\psi
}_{j}^{n}}\right)
\label{5.3}
\end{equation}

If we defined the conjugate momentum ${\pi}_{j}^{n}={i\psi}_{j}^{
\dag n}$ the Hamilton equation of motion are obtained variying the
Hamiltonian density ${\cal H}_j^n$ with respect to $i\psi_j^{\dag n}$ , that is
to say:
\begin{equation}
{1 \over {\tau}}{\Delta}_{n}{\tilde{\Delta}}_{\mit
j}{\psi}_{j}^{n}={1 \over i}{\tilde{\Delta}}_{\mit
n}\left\{{{\gamma}_{4}{\gamma}_{1}{1 \over {\mit
\varepsilon}}{\Delta}_{j}{\psi}_{j}^{n}+M
{\gamma}_{4}{\tilde{\Delta}}_{j}{\psi
}_{j}^{n}}\right\}
\label{5.4}
\end{equation}
from which it can be derived the Dirac equation on the lattice:
\begin{equation}
\left({{\gamma}_{1}{1 \over \varepsilon}{\Delta}_{j}
{\tilde{\Delta}}_{n}-{i\gamma}_{4}{1 \over \tau}
{\Delta}_{n}{\tilde{\Delta}}_{j}+M{\tilde{\Delta
}}_{j}{\tilde{\Delta}}_{n}}\right){\psi}_{j}^{n}
=0
\label{5.5}
\end{equation}

The same result was obtained by Bender, Milton and Sharp (Reference 1, formula
18) applying the method of finite elements to the action.

From (\ref{2.18}) and (\ref{5.5}) one proves for the Dirac Hamiltonian
(\ref{5.1})
\begin{equation}
{\Delta}_{n}{H}_{n}=0,\quad \mbox{hence}
\quad {\tilde{\Delta}}_{n}{H}_{n}={H}_{n}
\label{5.6}
\end{equation}

If we impose periodicity conditions in the fields we can prove by ``integration
by parts'' of (\ref{5.1})that $H_n$ is an hermitian function.

Applying the operator ${\gamma}_{1}{1 \over \varepsilon}{\nabla
}_{j}{\tilde{\nabla}}_{n}-{i\gamma}_{4}{1 \over
\tau}{\nabla}_{n}{\tilde{\Delta}}_{j}-M
{\tilde{\nabla}}_{j}{\tilde{\nabla}}_{n}$ on both sides of
(\ref{5.4}) we recover the wave equation (\ref{3.1}) for $\psi_j^n$

Let us construct solutions to (\ref{5.5}) of the form
\begin{equation}
{\psi}_{j}^{n}=w\left({k,E}\right){f}_{\mit
j}^{n}\left({k,E}\right)
\label{5.7}
\end{equation}
where $f_j^n\left({k,E}\right)$ is given by (\ref{3.2}).

The four component spinors $w(k,E)$ must satisfy
\begin{equation}
\left({i{\gamma}_{1}k-{\gamma}_{4}E
+M}\right)w\left({k,E}\right)=0
\label{5.8}
\end{equation}

Multiplying this equation from the left by $\left({i{\gamma}_{
1}k-{\gamma}_{4}E-M}\right)$ we obtain the
``dispersion relation'' for the Dirac equation
\begin{equation}
{E}^{2}-{k}^{2}={M}^{2}
\label{5.9}
\end{equation}

Following standard procedure we can construct a complete set of
Dirac wave functions of the form
\begin{eqnarray}
& & {1 \over \sqrt {N\varepsilon}}\sqrt {{M \over {E}_{m}}}{u}_{m}
{f}_{j}^{n}\left({{k}_{m},{E}_{m}}\right)
\label{5.10} \\ \noalign{\medskip}
& & {1 \over \sqrt {N\varepsilon}}\sqrt {{M \over {E}_{m}}}{v}_{m}
{f}_{j}^{*n}\left({{k}_{m},{E}_{m}}\right)
\label{5.11}
\end{eqnarray}
where
\begin{equation}
{E}_{m}=+\sqrt {{k}_{m}^{2}+{M}^{2}}, \quad {\mit
k}_{m}={2 \over \varepsilon}tan{\pi \over N}m
, \quad m=0,\pm 1,\ldots 
\label{5.12}
\end{equation}
and the spinors $u_m$ and $v_m$ are defined as usual:
\begin{eqnarray}
u_m &=&w\left({{k}_{m},{E}_{m}}\right)={\left({{{E}_{m}+M \over 2M}}\right)}^{{1 \over
2}}\left(\begin{array}{c}
1\\ \noalign{\smallskip}
\frac{k_m}{E_m+M}
\end{array}\right)
\label{5.13} \\ \noalign{\medskip}
{v}_{m}&=&w\left({{-k}_{m},{-E}_{m}}\right)
={\left({{{E}_{m}+M \over 2M}}\right)}^{{1 \over
2}}\left(\begin{array}{c}
\frac{k_m}{E_m+M}\\ \noalign{\smallskip}
1
\end{array}\right)
\label{5.14}
\end{eqnarray}

With the aid of this orthonormal set we can expand the fields in the usual way
(remenber \ref{3.21})
\begin{eqnarray}
{{\tilde{\Delta}}_{j}\psi}_{j}^{n}&=&{1 \over \sqrt
{N\varepsilon}}\sum\nolimits\limits_{m=-N/2}^{\mit
N/2-1} \sqrt {{M \over {E}_{m}}}\left({{u}_{m}{\mit
c}_{m}{f}_{j}^{n}\left({{k}_{m},{E}_{m}}\right)
+{v}_{m}{d}_{m}^{\dag}{f}_{j}^{*n}
\left({{k}_{m},{E}_{m}}\right)}\right)
\label{5.15} \\
{{\tilde{\Delta}}_{j}\overline{\psi}}_{j}^{n}&=&{1 \over
\sqrt {N\varepsilon}}\sum\nolimits\limits_{m=-N/
2}^{N/2-1} \sqrt {{M \over {E}_{m}}}\left({{\overline{\mit
u}}_{m}{c}_{m}^{\dag}{f}_{j}^{*n}
\left({{k}_{m},{E}_{m}}\right)+{\overline{v}}_{m}
{d}_{m}{f}_{j}^{n}\left({{k}_{m},{\mit
E}_{m}}\right)}\right)
\label{5.16}
\end{eqnarray}
where ${\overline{\psi}}_{j}^{n}\equiv {\psi}_{j}^{\dag \mit
n}{\gamma}_{4}$. Inverting this expressions, we find
\begin{eqnarray}
{c}_{m} &=& {1 \over \sqrt {N\varepsilon}}\sqrt {{M \over
{E}_{m}}}\varepsilon \sum\nolimits\limits_{j=0}^{\mit
N-1} {f}_{j}^{*n}\left({{k}_{m},{\mit
E}_{m}}\right){\overline{u}}_{m}{\gamma}_{
4}{\tilde{\Delta}}_{j}{\psi}_{j}^{n}
\label{5.17} \\
{c}_{m}^{\dag}&=&{1 \over \sqrt {N\varepsilon}}\sqrt {{M
\over {E}_{m}}}\varepsilon \sum\nolimits\limits_{j
=0}^{N-1} {\tilde{\Delta}}_{j}{\overline{\psi
}}_{j}^{n}{\gamma}_{4}{u}_{m}{f}_{j}^{n}
\left({{k}_{m},{E}_{m}}\right)
\label{5.18} \\
{d}_{m}&=&{1 \over \sqrt {N\varepsilon}}\sqrt {{M \over
{E}_{m}}}\varepsilon \sum\nolimits\limits_{j=0}^{\mit
N-1} {\tilde{\Delta}}_{j}{\overline{\psi}}_{j}^{n}{\mit
\gamma}_{4}{v}_{m}{f}_{j}^{*n}\left({{\mit
k}_{m},{E}_{m}}\right)
\label{5.19} \\
{d}_{m}^{\dag}&=&{1 \over \sqrt {N\varepsilon}}\sqrt {{M
\over {E}_{m}}}\varepsilon \sum\nolimits\limits_{j
=0}^{N-1} {f}_{j}^{n}\left({{k}_{m},{\mit
E}_{m}}\right){\overline{v}}_{m}{\gamma}_{
4}{\tilde{\Delta}}_{j}{\psi}_{j}^{n}
\label{5.20}
\end{eqnarray}

For the quantization of the Dirac field we start from the Hamilton equation of
motion (\ref{5.4}) where now ${\psi}_{\alpha j}^{n}$ and ${\psi}_{\alpha
j}^{\dag n}$ are in general non commuting operators. Using the same
technique as in the Klein-Gordon field we can prove that the equal time
anticommutation relations hold for any $n$ 
\begin{eqnarray}
{\left[{\tilde{\Delta}_j\psi_{\alpha j}^n, \tilde{\Delta}_{j'}\psi_{\beta
j'}^{\dag n}}\right]}_{+}&=&{1 \over \varepsilon}\delta_{\alpha \beta
}\delta_{jj'} \label{5.21} \\
{\left[{\tilde{\Delta}_j\psi_{\alpha j}^n,\tilde{\Delta}_{j'}\psi_{\beta
j'}^n}\right]}_{+}&=&{\left[{\tilde{\Delta}_j\psi_{\alpha j}^{\dag n},
\tilde{\Delta}_{j'}\psi_{\beta j'}^{\dag n}}\right]}_{+}=0 \label{5.22}
\end{eqnarray}
provided it holds for $n = 0$.

Introducing the Hamiltonian (\ref{5.1}) in the Heisenberg equation of motion and
using the anticommutation relations (\ref{5.21}, \ref{5.22}) we recover the
Dirac equation, namely,
\begin{equation}
{1 \over \tau} \Delta_n \left(\tilde{\Delta}_j \psi_j^n\right) ={1
\over i} \left[\tilde{\Delta}_n\left(\tilde{\Delta
}_j\psi_j^n\right),H_n\right]={1 \over i}\tilde{\Delta}_n\left[\tilde{\Delta
}_j\psi_j^n,H_n\right]={1 \over i}\tilde{\Delta}_n\left\{\gamma_4\gamma
_1{1 \over \varepsilon}\Delta_j\psi
_j^n+M\gamma_4\tilde{\Delta}_j\psi
_j^n\right\}
\label{5.23}
\end{equation}

Using (\ref{5.21}, \ref{5.22}) we can derived the anticommutation relations for
the creation and annihilation operators
\begin{equation}
{\left[{{c}_{m},{c}_{m'}^{\dag}}\right]}_{+}={\delta
}_{mm'} \qquad {\left[{{d}_{m},{d}_{m'}^{\dag
}}\right]}_{+}={\delta}_{mm'}
\label{5.24}
\end{equation}
with other anticommutation relations vanishing.

Finally, the Hamiltonian (\ref{5.1}) can be written in terms of these operators
\begin{equation}
H=\sum\nolimits\limits_{m=-N/2}^{N/2-1} {\mit
E}_{m}\left({{c}_{m}^{\dag}{c}_{m}+{d}_{m}^{
\dag}{d}_{m}-1}\right)
\label{5.25}
\end{equation}

Similarly the momentum and charge operators are written:
\begin{eqnarray}
P &=& -i\sum\nolimits\limits_{j=0}^{N-1}
\left({{\tilde{\Delta}}_{j}{\psi}_{j}^{\dag \mit
n}}\right)\left({{\Delta}_{j}{\psi}_{j}^{n}}\right)
=\sum\nolimits\limits_{m=-N/2}^{N/2-1} {\mit
k}_{m}\left({{c}_{m}^{\dag}{c}_{m}+{d}_{m}^{
\dag}{d}_{m}}\right)
\label{5.26} \\
Q &=& \sum\nolimits\limits_{j=0}^{N-1} \left({{\tilde{\Delta
}}_{j}{\psi}_{j}^{\dag n}}\right)
\left({{\tilde{\Delta}}_{j}{\psi}_{j}^{n}}\right)
=\sum\nolimits\limits_{m=-N/2}^{N/2-1} \left({{\mit
c}_{m}^{\dag}{c}_{m}-{d}_{m}^{\dag}{d}_{m}
}\right)+2N
\label{5.27}
\end{eqnarray}

The discrete analog of the Feynman propagator for the Dirac field is defined as
before
\begin{equation}
{S}_{\alpha \beta ,j-j'}^{n-n'}=\left \langle 0
\left\vert T{\psi}_{\alpha j}^{n}{\overline{\psi}}_{\beta
j'}^{n'}\right\vert 0\right\rangle
\label{5.28}
\end{equation}
where the time order operator acts as follows
\begin{equation}
T{\psi}_{j}^{n}{\overline{\psi}}_{j'}^{n
'}=\left\{
\begin{array}{cl}
\psi_j^n\overline{\psi}_{j'}^{n'} & \mbox{for} \quad n>n' \\
\noalign{\smallskip} -\overline{\psi}_{j'}^{n'}\psi_j^n & \mbox{for}\quad n'>n
\end{array}
\right.
\label{5.29}
\end{equation}

Following standard methods we find:
\begin{eqnarray}
{S}_{\alpha \beta ,j-j'}^{n-n'}&=&{\mit
\vartheta}_{n}{1 \over N\varepsilon}\sum\nolimits\limits_{\mit
m=-N/2}^{N/2-1} {{\left({1-{i \over 2}\mit
\varepsilon {k}_{m}}\right)}^{2} \over 2i{E}_{m}}
{\left({{\gamma}_{1}{k}_{m}+i{E}_{m}+\mit
iM}\right)}_{\alpha \beta}{f}_{j-j'}^{n-\mit
n'}\left({{k}_{m},{E}_{m}}\right) \nonumber \\
&-&{\vartheta}_{-n}{1 \over N\varepsilon}
\sum\nolimits\limits_{m=-N/2}^{N/2-1} {{\left({1-{1
\over 2}i\varepsilon {k}_{m}}\right)}^{2} \over 2\mit
i{E}_{m}}{\left({{\gamma}_{1}{k}_{m}+i{E}_{m}
+iM}\right)}_{\alpha \beta}{f}_{j-j'}^{*\mit
n-n'}\left({{k}_{m},{E}_{m}}\right)
\label{5.30}
\end{eqnarray}

We can also derive the Feynman propagator from the Dirac equation. Taking the
Fourier transform of (\ref{5.5}) we get
\begin{equation}
{\hat{G}}_{m}^{r}={\left({1-{1 \over 2}i\varepsilon {\mit
k}_{m}}\right)\left({1+{1 \over 2}i\tau {E}_{r}}\right) \over
i{\gamma}_{1}{k}_{m}-{\gamma}_{4}{E}_{r}+\mit
M}=\left({1-{1 \over 2}i\varepsilon {k}_{m}}\right)
\left({1+{1 \over 2}i\tau {E}_{r}}\right){{{i\gamma
}_{1}{k}_{m}-{\gamma}_{4}{E}_{r}-M \over
{E}_{r}^{2}-{k}_{m}^{2}-{M}^{2}}}
\label{5.31}
\end{equation}
hence
\begin{equation}
{G}_{j-j'}^{n-n'}={1 \over N\varepsilon 
N\tau}\sum\nolimits\limits_{m=-N/2}^{N/
2-1} \sum\nolimits\limits_{r=-N/2}^{N/2-1}
{\hat{G}}_{m}^{r}{f}_{j-j'}^{n-n
'}\left({{k}_{m},{E}_{r}}\right)
\label{5.32}
\end{equation}
Both function (\ref{5.30}) and (\ref{5.32}) are the Green function for the Dirac
equations, therefore they are equal up to a constant.

Our model for the fermion field satisfies the following conditions:

\begin{itemize}
\item[i)] the hamiltonian (\ref{5.1}) is traslational invariant with respect to
the space indices.
\item[ii)] the hamiltonian is hermitian.
\item[iii)] for $M = 0$, the wave equation (\ref{5.5}) is invariant under global
chiral transformations.
\item[iv)] there is no ``fermion doubling'' as it can be seen in (\ref{5.12}). In
fact, $E_m$ takes the value $M$ at $m = 0$ and nowhere else.
\item[v)] the hamiltonian is non-local. Using the finite Fourier transform
(\ref{2.11}) the hamiltonian $H\left({{k}_{m}}\right)
=i{\gamma}_{4}{\gamma}_{1}{k}_{m}+{\mit
\gamma}_{4}M$, given by (\ref{5.8}), and consequently the
dispersion relations (\ref{5.12}) are smooth functions of $k_m$, except for $m =
N/2$. Therefore our model escapes the no-go theorem of Nielsen and
Ninomiya \cite{Niel}. \end{itemize}

\section{Anomalies in axial vector current} 
In order to construct vector and axial currents in terms of the fermions fields
in the presence of external electromagnetic vector potential, we multiply
(\ref{5.5}) from the left by ${\tilde{\Delta}}_{j}{\tilde{\Delta
}}_{n}{\overline{\psi}}_{j}^{n}$ and then we multiply the
adjoint equation of (\ref{5.5}) from the right by ${\tilde{\Delta
}}_{j}{\tilde{\Delta}}_{n}{\psi}_{j}^{n}$. Adding together both
results and using (\ref{2.18}) we find:
\begin{equation}
{1 \over \varepsilon}{\Delta}_{j}\left({{\tilde{\Delta
}}_{n}{\overline{\psi}}_{j}^{n}{\gamma}_{
1}{\tilde{\Delta}}_{n}{\psi}_{j}^{n}}\right)-{1 \over \mit
\tau}{i\Delta}_{n}\left({{\tilde{\Delta}}_{j}
{\overline{\psi}}_{j}^{n}{\gamma}_{4}{\tilde{\Delta
}}_{j}{\psi}_{j}^{n}}\right)=0
\label{6.1}
\end{equation}

This equation can be considered the discrete version of the ``conservation law''
for the vector current
\begin{equation}
{j}_{1}=i\left({{\tilde{\Delta}}_{n}{\overline{\psi
}}_{j}^{n}}\right){\gamma}_{1}\left({{\tilde{\Delta}}_{\mit
n}{\psi}_{j}^{n}}\right), \quad {j}_{4}=i
\left({{\tilde{\Delta}}_{j}{\overline{\psi}}_{j}^{n}}\right)
{\gamma}_{4}\left({{\tilde{\Delta}}_{j}{\psi
}_{j}^{n}}\right)
\label{6.2}
\end{equation}

The same equation (\ref{6.1}) can be applied to the axial current
\begin{equation}
{j}_{1}^{5}=i\left({{\tilde{\Delta}}_{n}{\overline{\mit
\psi}}_{j}^{n}}\right){\gamma}_{1}{\gamma}_{
5}\left({{\tilde{\Delta}}_{n}{\psi}_{j}^{n}}\right), \quad
{j}_{4}^{5}=i\left({{\tilde{\Delta}}_{j}
{\overline{\psi}}_{j}^{n}}\right){\gamma}_{4}{\gamma
}_{5}\left({{\tilde{\Delta}}_{j}{\psi}_{j}^{n}}\right)
\label{6.3} \end{equation}

Both currents are invariant under global chiral transformations but they are
not invariant under $U(1)$-gauge transformations:
\begin{equation}
{\psi}_{j}^{n}\rightarrow {\Omega}_{j}^{n}{\psi
}_{j}^{n}, \quad {\overline{\psi}}_{j}^{n}\rightarrow
{\overline{\psi}}_{j}^{n}{\Omega}_{j}^{n}
\label{6.4}
\end{equation}
with ${\Omega}_{j}^{n}$ some unitary function of discrete variables. In
order to have gauge invariance we define a gauge field on the lattice
\begin{equation}
{U}_{j}^{nn'}={1+{i \over 4}\tau \left\{{{\left({{\mit
A}_{0}}\right)}_{j}^{n}+{\left({{A}_{0}}\right)}_{\mit
j}^{n'}}\right\} \over 1-{i \over 4}\tau 
\left\{{{\left({{A}_{0}}\right)}_{j}^{n}+{\left({{\mit
A}_{0}}\right)}_{j}^{n'}}\right\}}
\label{6.5}
\end{equation}
and
\begin{equation}
{U}_{jj'}^{n}={1+{i \over 4}\varepsilon 
\left\{{{\left({{A}_{1}}\right)}_{j}^{n}+{\left({{\mit
A}_{1}}\right)}_{j'}^{n}}\right\} \over 1-{i \over
4}\varepsilon \left\{{{\left({{A}_{1}}\right)}_{\mit
j}^{n}+{\left({{A}_{1}}\right)}_{j'}^{n}}\right\}}
\label{6.6}
\end{equation}
where $(A_1, iA_0)$ are the two component electromagnetic vector potential, each
of them satisfying the wave equation (\ref{3.1}) with $M = 0$ namely, 
\begin{equation}
\left({{1 \over \tau^2}{\nabla}_{n} {\Delta}_n {\tilde{\nabla}}_j
{\tilde{\Delta}}_j - {1 \over {{\varepsilon}^2}}{\nabla}_j
\Delta_j {\tilde{\nabla}}_n {\tilde{\Delta}}_n}
\right){\left({{A}_{1}}\right)}_{\mit j}^{n}=0 
\label{6.7} 
\end{equation}
and similarly for ${\left({{A}_{0}}\right)}_{j}^{n}$.

The gauge fields $U_j^{nn'}$ are associated with the link between the points 
$\left({j,n}\right)\rightarrow \left({j,n
'}\right)$ in the positive direction, and they transform under the gauge group
as follows:
\begin{eqnarray}
{U}_{j}^{nn'}&\rightarrow & {\Omega}_{j}^{\dag n}{\mit
U}_{j}^{nn'}{\Omega}_{j}^{n'}
\label{6.8} \\
{U}_{jj'}^{n}&\rightarrow & {\Omega}_{j}^{\dag n}
{U}_{jj'}^{n}{\Omega}_{j'}^{n}
\label{6.9}
\end{eqnarray}

Inserting these fields in the vector and axial currents between fermion fields
at separated points, we get
\begin{eqnarray}
{j}_{1}&=&{i \over 4}\left({{\overline{\psi}}_{j}^{n}
{\gamma}_{1}{\psi}_{j}^{n}+{\overline{\psi
}}_{j}^{n}{\gamma}_{1}{U}_{j}^{n,n+1}{\psi
}_{j}^{n+1}+{\overline{\psi}}_{j}^{n+1}{\gamma}_{1}{\mit
U}_{j}^{n+1,n}{\psi}_{j}^{n}+{\overline{\psi
}}_{j}^{n+1}{\gamma}_{1}{\psi}_{j}^{n+1}}\right)
\label{6.10} \\
{j}_{4}&=&{i \over 4}\left\{{{\overline{\psi}}_{j}^{n}
{\gamma}_{4}{\psi}_{j}^{n}+{\overline{\psi
}}_{j}^{n}{\gamma}_{4}{U}_{j,j+1}^{n}
{\psi}_{j+1}^{n}+{\overline{\psi}}_{j+1}^{\mit
n}{\gamma}_{4}{U}_{j+1,j}^{n}{\psi
}_{j}^{n}+{\overline{\psi}}_{j+1}^{n}{\gamma}_{
4}{\psi}_{j+1}^{n}}\right\}
\label{6.11}
\end{eqnarray}
and similarly for $j_1^5$ and $j_4^5$. Using (\ref{6.8}) and (\ref{6.9}) we can
prove that all these expresions for the vector and axial currents are invariant
under the gauge transformations (\ref{6.4}).

Now we want to calculate the vacuum expectation value of the divergence of the
vector and axial current. We assume that this vacuum expectation value
approches the corresponding non interacting fields for vanishing spacial
separation of the fields\cite{Brow}. Thus we use solution of the free massless
fermion field (\ref{5.5}) with $M=0$, namely,
\begin{equation}
{\psi}_{j}^{n}={1 \over \sqrt {N\varepsilon}}
\sum\nolimits\limits_{m=-{N \over 2}}^{{N \over 
2}-1} \left({{u}_{m}{c}_{m}{f}_{j}^{n}\left({{k}_{m}
,{\omega}_{m}}\right)+{v}_{m}{d}_{m}^{\dag
}{f}_{j}^{n}\left({{k}_{m},-{\omega}_{m}}\right)}\right)
\label{6.12}
\end{equation}
with ${u}_{m}=\left({\matrix{1\cr
1\cr}}\right)$, and ${v}_{m}=\left({\begin{array}{c}
1\\
-1
\end{array}
}\right)$, and the operators $c_m$ and $d_m$ satisfying the usual
anticommutation relations:
\begin{equation}
{\left[{c_m,c_{m'}^{\dag}}\right]}_{+}=\delta_{mm'} \quad,\quad
{\left[{d_m,d_{m'}^{\dag}}\right]}_{+}=\delta_{mm'}
\label{6.13}
\end{equation}
Applying these operators to the vacuum of the Fock space, we get 
\begin{eqnarray}
{d}_{m}^{\dag}\vert 0\rangle &=& {c}_{m}^{\dag}\vert
0\rangle =0 \mbox{for} \quad m \ge 0
\label{6.14} \\
{d}_{m}\vert 0\rangle &=&{c}_{m}\vert 0\rangle =0 \mbox{for}
\quad m <0
\label{6.15}
\end{eqnarray}

Collecting these properties we obtain for the axial current
\begin{eqnarray*}
\lefteqn{
\left\langle 0\left\vert{1 \over \varepsilon}{\Delta}_{j}{\mit
j}_{1}^{5}-{1 \over \tau}i{\Delta}_{n}{\mit
j}_{4}^{5}\right\vert 0\right\rangle=}
\\ \noalign{\smallskip}
& &{i \over 4\varepsilon}\left\langle 0\left\vert{\overline{\psi}}_{j
+1}^{n}{\gamma}_{1}{\gamma}_{5}{\psi}_{j
+1}^{n+1}\right\vert 0\right\rangle {U}_{j+1}^{n,n
+1}-{i \over 4\varepsilon}\left\langle 0\left\vert {\overline{\psi
}}_{j}^{n}{\gamma}_{1}{\gamma}_{5}{\psi}_{j}^{n
+1}\right\vert 0\right\rangle {U}_{j}^{n,n+1}
\\ \noalign{\smallskip}
&+&{{i}^{2} \over 4\tau}\left\langle 0\left\vert{\overline{\psi
}}_{j}^{n+1}{\gamma}_{4}{\gamma}_{5}{\psi}_{j
+1}^{n+1}\right\vert 0\right\rangle {U}_{j,j+1}^{n+1}-{{\mit
i}^{2} \over 4\tau}\left\langle 0\left\vert{\overline{\psi}}_{j}^{n}
{\gamma}_{4}{\gamma}_{5}{\psi}_{j+1}^{n}
\right\vert 0\right\rangle {U}_{j,j+1}^{n}+\mbox{c.c.} =
\\ \noalign{\smallskip}
&-&{1 \over N\varepsilon}\left\{{\sum\nolimits\limits_{\mit
m=0}^{N/2-1} \left({{1-{i \over 2}\tau {\omega}_{m}
\over 1+{i \over 2}\tau {\omega}_{m}}}\right)
+\sum\nolimits\limits_{m=-N/2}^{-1} \left({{1+{i \over
2}\tau {\omega}_{m} \over 1-{i \over 2}\tau {\omega
}_{m}}}\right)}\right\}{1 \over 4\varepsilon}\left({{U}_{j
+1}^{n,n+1}-{U}_{j}^{n,n+1}}\right)
\\ \noalign{\smallskip}
&+&{1\over N\varepsilon}\left\{{-\sum\nolimits\limits_{m
=0}^{N/2-1} \left({{1+{i \over 2}\varepsilon {k}_{m}
\over 1-{i \over 2}\varepsilon {k}_{m}}}\right)
+\sum\nolimits\limits_{m=-N/2}^{-1} \left({{1+{i \over
2}\varepsilon {k}_{m} \over 1-{i \over 2}\varepsilon
{k}_{m}}}\right)}\right\}{1 \over 4\tau}\left({{U}_{j
,j+1}^{n+1}-{U}_{j,j+1}^{n}}\right)
+\mbox{c.c.} =
\\ \noalign{\smallskip}
& &{-1 \over 2Ntg{\pi \over N}}{{1 \over \mit
\varepsilon}{\tilde{\Delta}}_{n}{\Delta}_{j}
{\left({{A}_{0}}\right)}_{j}^{n} \over 1-i\varepsilon 
{\tilde{\Delta}}_{j}{\tilde{\Delta}}_{n}{\left({{\mit
A}_{0}}\right)}_{j}^{n}-{{\varepsilon}^{2} \over
4}{\tilde{\Delta}}_{j}^{2}{\left({{A}_{0}}\right)}_{2\mit
j}^{2n}} + {{e}^{-i\pi /N} \over 
2Nsen{\pi \over N}}{{1 \over \tau}{\tilde{\Delta
}}_{j}{\Delta}_{n}{\left({{A}_{1}}\right)}_{\mit
j}^{n} \over 1-i\varepsilon {\tilde{\Delta}}_{j}
{\tilde{\Delta}}_{n}{\left({{A}_{1}}\right)}_{j}^{n}
-{{\varepsilon}^{2} \over 4}{\tilde{\Delta}}_{n}^{
2}{\left({{A}_{1}}\right)}_{2j}^{2n}}+\mbox{c.c.}
\end{eqnarray*}
which in the limit, $N\rightarrow \infty ,\quad \varepsilon 
\rightarrow 0,\quad \tau \rightarrow 0,$ becomes
\begin{equation}
\left\langle 0 \left\vert{\partial}_{1}{j}_{1}^{5}+{\partial}_{4}{j}_{
4}^{5}\right\vert 0 \right\rangle =-{1 \over \pi}\left({{\partial}_{1}{A}_{
0}-{\partial}_{0}{A}_{1}}\right)=-{1 \over \pi}{\mit
F}_{10}
\label{6.16}
\end{equation}
as required. Similarly we can prove
$$\left\langle 0 \left\vert{1 \over \varepsilon}{\Delta}_{j}{\mit
j}_{1}+{1 \over \tau}i{\Delta}_{n}{\mit
j}_{4}\right\vert 0 \right\rangle =0$$

We come to the conclusion that our model leads to an interaction which is
$U(1)$-gauge invariant, but the divergence of the axial current gives in the
continuous limit a photon mass, as expected by the axial anomaly.

For the sake of simplicity in the notation we have worked all the formulas in
$(1+1)$-dimension. The expressions in this paper can be easily generalized to
$(3+1)$-dimensional lattice by the use of the promediate fields:
$${\tilde{\nabla}}_{{j}_{1}}{\tilde{\Delta}}_{{j}_{
1}}{\tilde{\nabla}}_{{j}_{2}}{\tilde{\Delta}}_{{j}_{
2}}{\tilde{\nabla}}_{{j}_{3}}{\tilde{\Delta}}_{{j}_{3}}\mit
\phi \left({{\varepsilon}_{1}{j}_{1},{\varepsilon
}_{2}{j}_{2},{\varepsilon}_{3}{j}_{3},\tau
n}\right)\rightarrow \phi \left({{x}_{1},{x}_{
2},{x}_{3},t}\right)$$
and similary for space an time differences.

We have discussed symmetries and conservation laws for this scheme
elsewhere\cite{Lor2}.

\section{Acknowledgments} 
This work has been partially supported by the Vicerrectorado de Investigaci\'on
of the University of Oviedo, and by Volkswagen Foundation.

The author wants to thanks Professors L. V\'azquez, L.A. Fern\'andez, J. Mateos
and A.Gz. Arroyo for useful conversations. He also thanks Prof. H.B. Doebner and
members of the Arnold Sommerfeld Institut at the University of Clausthal, where
part of this work was finished, for hospitality and advise.

\end{document}